\documentclass[letterpaper, 10 pt, conference]{ieeeconf}  

\IEEEoverridecommandlockouts                              

\usepackage{cite}
\usepackage{amsmath,amssymb,amsfonts}
\usepackage{algorithmic}
\usepackage{graphicx}
\usepackage{subfig}    
\usepackage{float}      
\usepackage{lipsum}     
\usepackage{lineno}
\usepackage{tabularx}
\usepackage{setspace}

\usepackage[ruled,linesnumbered]{algorithm2e}
\SetKwRepeat{Do}{do}{while}
\SetKw{KwInit}{Initialization:}

\usepackage{booktabs}
\usepackage{textcomp}
\usepackage{xcolor}

\usepackage{balance}  

\title{\LARGE \bf
A Residual Variance Matching Recursive Least Squares Filter \\ for Real-time UAV Terrain Following
}

\author{Xiaobo Wu$^{1}$, Youmin Zhang$^{1*}$
	\thanks{$^{1}$Xiaobo Wu, Youmin Zhang are with the Department of Mechanical, Industrial and Aerospace Engineering, Concordia University, Montreal, Quebec H3G 1M8, Canada (wu\_xiaob@live.concordia.ca, youmin.zhang@concordia.ca)
	}%
}

\begin{document}
\maketitle
\thispagestyle{empty}
\pagestyle{empty}

\begin{abstract}
Accurate real-time waypoints estimation for the UAV-based online Terrain Following during wildfire patrol missions is critical to ensuring flight safety and enabling wildfire detection.
However, existing real-time filtering algorithms struggle to maintain accurate waypoints under measurement noise in nonlinear and time-varying systems, posing risks of flight instability and missed wildfire detections during UAV-based terrain following.
To address this issue, a Residual Variance Matching Recursive Least Squares (RVM-RLS) filter, guided by a Residual Variance Matching Estimation (RVME) criterion, is proposed to adaptively estimate the real-time waypoints of nonlinear, time-varying UAV-based terrain following systems.
The proposed method is validated using a UAV-based online terrain following system within a simulated terrain environment.
Experimental results show that the RVM-RLS filter improves waypoints estimation accuracy by approximately 88$\%$ compared with benchmark algorithms across multiple evaluation metrics.
These findings demonstrate both the methodological advances in real-time filtering and the practical potential of the RVM-RLS filter for UAV-based online wildfire patrol.
\end{abstract}


\section{INTRODUCTION}

\begin{figure}[h]
	\centering
	\includegraphics[width=3.2in]{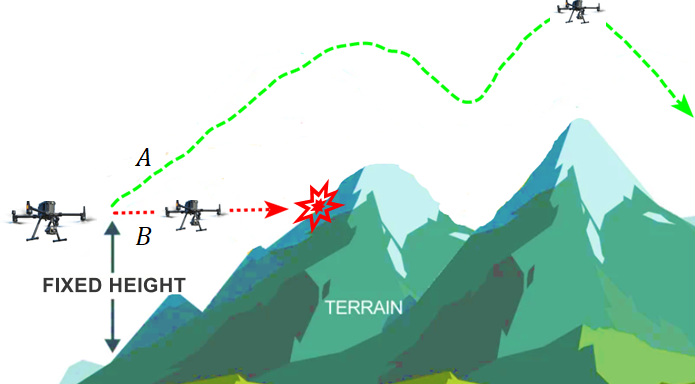}
	\caption{Longitudinal path comparison of UAV-based online terrain following in wildfire patrol: the dashed green line $A$ denotes the terrain following strategy, and the dotted red line $B$ represents the fixed altitude approach.}
	\label{fig2_comparasion}
\end{figure}

Wildfires inflict substantial devastation and damage \cite{kolden2024wildfires, jones2024state}.
With the rapid development of unmanned aerial vehicles (UAVs), UAV-based wildfire patrol have become crucial for early detection and rapid emergency response to prevent catastrophic losses \cite{qiao2024early, zhu2025multiscale}. 
The UAV-based online terrain following technology \cite{campos2016height, de2024analise} plays on important role in wildfire patrol by ensuring consistent data acquisition accuracy and prevents potential collisions between the UAV and undulating terrain \cite{samar2011autonomous, singh2023high}.
Fig.~\ref {fig2_comparasion} demonstrates the longitudinal path comparison between terrain following strategy and fixed altitude approach. 
By leveraging onboard sensors and computing units, online terrain following systems enable real-time terrain perception and path planning. However, sensor measurement error in unknown environments inevitably introduce outliers and randon noise, which can degrade the accuracy of real-time data collection, thereby reducing the precision of waypoints of the terrain following, and in severe cases, threaten flight safety.

Real-time filtering plays a vital role in mitigating these effects by removing noise, estimating parameters, and predicting states as signals evolve. 
The Least Mean Squares (LMS) algorithm is widely favored for its simplicity and high computational efficiency. However, it suffers from slow convergence and high sensitivity to the step-size selection. Its variants, such as the Normalized LMS, retain LMS’s low complexity while improving convergence stability.
The Recursive Least Squares (RLS) \cite{pan2022hierarchical} algorithm achieves significant improvements in both convergence speed and steady-state accuracy, but at the cost of increased computational complexity and memory requirements. Furthermore, its adaptiveness and stability are constrained by the choice of the forgetting factor. Variants of RLS attempt to alleviate these issues by introducing dynamic forgetting factors and improving robustness under uncertainty.
Both LMS and RLS algorithms rely on online parameter identification and prediction based on input-output signals for adaptive filtering.

The Kalman Filter (KF) \cite{wang2023kalman} serves as an optimal linear estimator and is widely applied to dynamic system state estimation. When the system model becomes nonlinear or non-Gaussian, the Extended Kalman Filter (EKF) \cite{mossaddek2025efficient}, Unscented Kalman Filter (UKF) \cite{bucci2022evaluation}, and Particle Filter (PF) \cite{wang2017survey} are typically employed. However, KF-based algorithms are highly dependent on accurate state-transition models. 

In the absence of such models or when they are inaccurate, the filter’s performance often deteriorates to that of RLS.
Recently, machine learning (ML) and neural networks (NNs)-based filtering approaches, such as the Kalman Filter with Long Short-Term Memory (KF-LSTM) \cite{tian2024application} and Deep Neural Network Kalman Filter (DNN-KF) \cite{verma2025hybrid} have shown strong capability in handling non-Gaussian noise, nonlinear dynamics, and time-varying models. Nevertheless, these methods still face challenges in real-time processing, data dependency, and model training complexity, limiting their practical deployment in fast-evolving scenarios such as UAV wildfire patrols.

Although significant progress has been made in real-time filtering algorithms, existing approaches still struggle to maintain a balance among accuracy, robustness, and adaptability.
Consequently, there is a critical need for a real-time filtering strategy that can achieve high estimation accuracy, strong robustness, and adaptability to nonlinear and time-varying systems.

To address this challenge, the Residual Variance Matching–Recursive Least Squares (RVM-RLS) filter is proposed.
The main contributions of this work are summarized as follows:
\begin{enumerate}
    \item
    We proposed the RVM-RLS filter by introducing the new Residual Variance Matching Estimation (RVME) principle.
    It integrates adaptive learning, nonlinear modeling, and robustness enhancement within a dual-recursive estimation framework to achieve accurate real-time waypoints estimation through a data-driven approach for UAV-based online terrain following systems with nonlinear and time-varying characteristics.
    \item 
     A real-time UAV terrain following system equipped with a LiDAR rangefinder is designed and modeled to validate the proposed filter.
    \item
    Simulations are conducted independently and comparatively in a simulated terrain to evaluate the performance of the proposed filter against representative real-time filtering algorithms across multiple evaluation metrics.
\end{enumerate}

The RVM-RLS filter is introduced step by step in Section \uppercase\expandafter{\romannumeral2}.
The UAV-based online terrain following system is designed and modeled to generate real-time waypoints in Section \uppercase\expandafter{\romannumeral3}.
The simulation and results are implemented in detail in Section \uppercase\expandafter{\romannumeral4}. 
Finally, Section
\uppercase\expandafter{\romannumeral5} 
summarizes the paper and outlines the future work.

\section{RVM-RLS Filter}
This section begins with the construction of the RLS framework based on a polynomial regression model[] and the $3\sigma$ rule [], followed by the design of the proposed RVM-RLS filter, and ends with the validation of the equivalence between the MMSE estimator [] and the RVME principle.

\subsection{RLS framework with the polynomial regression model and the 3$\sigma$ rule}
The $3\sigma$ rule is commonly used as a heuristic threshold for outlier detection. The polynomial regression model, a form of regression analysis in machine learning, is effective in modeling nonlinear relationships. Moreover, the Recursive Least Squares (RLS) filter provides real-time parameter estimation capability. Therefore, integrating the PR model and the $3\sigma$ rule into the RLS framework enables robust real-time parameter estimation for nonlinear systems.

For a nonlinear system subject to disturbances and noise, at time instant $i$, the measurement $y_{(i)}$ is modeled by an $m$-degree PR model to capture nonlinear trends, with time $t_{(i)}$ as the independent variable and $y_{(i)}$ as the dependent variable, leading to the following fitted equation: \eqref{RLS_SYS}.
\begin{equation}
	\label{RLS_SYS}
	\begin{split}
y_{(i)}={\mathbf{\Phi}_{(i)}^\mathsf{T}\boldsymbol{\theta}_{(i)}}+v_{(i)}, (i=1,2,\ldots,n)
	\end{split}
\end{equation}
where, $\boldsymbol{\theta}_{(i)}$ is the parameter vector, $\mathbf{\Phi}_{(i)}^\mathsf{T}=[1 \quad t_{(i)}^{}\quad t_{(i)}^{2}\quad \ldots \quad t_{(i)}^{m}]$ is the polynomial basis function about time $i$, and $m$ is the degree of the polynomial. 
The noise term $v_{(i)}$ is assumed to follow a Gaussian distribution $v_{(i)} \sim \mathcal{N}(0,\sigma^2)$.

After initializing $\hat{\boldsymbol{\theta}}_{(i)}$ (the estimation of the parameter vector) and $\mathbf{P}_{(i)}$ (inverse autocorrelation matrix), based on $\mathbf{\Phi}_{(i+1)}^\mathsf{T}$ and $y_{(i+1)}$, the proposed RLS framework can recursively obtain the updated estimation of the parameter vector $\hat{\boldsymbol{\theta}}_{(i+1)}$ and the prediction of the measurement $\hat{y}_{(i+1)}$.\\
\textbf{Main Loop of the Proposed RLS Framework} \\
\begin{enumerate}
    \item Predict the measurement (state) $\hat{y}_{(i+1)}$:
\begin{equation}
	\label{RLS_pre}
	\begin{split}
    \hat{y}_{(i+1)} = {\mathbf{\Phi}_{(i+1)}^\mathsf{T} \hat{\boldsymbol{\theta}}}_{(i)}
	\end{split}
\end{equation}
    \item Compute residual of measurement $y_{(i+1)}$:
\begin{equation}
	\label{RLS_ERRO}
	\begin{split}
    r_{(i+1)}= y_{(i+1)}-\hat{y}_{(i+1)}
	\end{split}
\end{equation}
\item Outlier detection based on 3$\sigma$ rule
    \eqref{RLS_ERROR_1}:
\begin{equation}
	\label{RLS_ERROR_1}
	\begin{split}
    r_{(i+1)} = 0,  \quad if ({|r_{(i+1)}|>3}\sigma_{v} )
	\end{split}
\end{equation}
    \item Compute the gain vector \eqref{eq:RLS_GAIN}:
\begin{equation}
	\label{eq:RLS_GAIN}
	\begin{split}
\mathbf{K}_{(i+1)}=\frac{\mathbf{P}_{(i)}\mathbf{\Phi}_{(i+1)}}{\lambda+\mathbf{\Phi}_{(i+1)}^\mathsf{T}\mathbf{P}_{(i)}\mathbf{\Phi}_{(i+1)}}
	\end{split}
\end{equation}
    \item Update the estimation of the parameter vector \eqref{RLS_Para_Updat}:
    
\begin{equation}
	\label{RLS_Para_Updat}
	\begin{split}
\hat{\boldsymbol{\theta}}_{(i+1)}=\hat{\boldsymbol{\theta}}_{(i)}+\mathbf{K}_{(i+1)}r_{(i+1)}
	\end{split}
\end{equation}
    \item Update the inverse autocorrelation matrix
            \eqref{RLS_coo_matrix}:
\begin{equation}
	\label{RLS_coo_matrix}
	\begin{split}
\mathbf{P}_{(i+1)}=\frac{1}{\lambda}\left(\mathbf{P}_{(i)}-\mathbf{K}_{(i+1)}\mathbf{\Phi}_{(i+1)}^\mathsf{T}\mathbf{P}_{(i)}\right)
	\end{split}
\end{equation}
\end{enumerate}
where, $\boldsymbol{\hat{\theta}}_{(i+1)}\in\mathbb{R}^{m+1}$, $\mathbf{t}_{(i+1)}\in\mathbb{R}^{m+1}$, $\mathbf{K}_{(i+1)}\in\mathbb{R}^{m+1}$ and $\mathbf{P}_{(i+1)}\in\mathbb{R}^{(m+1)\times(m+1)}$, $(m+1)$ is the number of estimated parameters of $\boldsymbol{\hat{\theta}}_{(i+1)}$,  $\lambda \in (0,\,1]$  is the static forgetting factor. 
A static forgetting factor plays a limited role in balancing adaptability and stability in time-varying systems. 
To enhance tracking capability while maintaining the smoothness to noise for a dynamic system, the RVM-RLS filter is further proposed by adding a forgetting factor estimator into the proposed RLS framework based on RVME paradigm.

\subsection{RVM-RLS Filter}

\begin{figure}[h!t]
    \centering
    \includegraphics[width=3.5in]{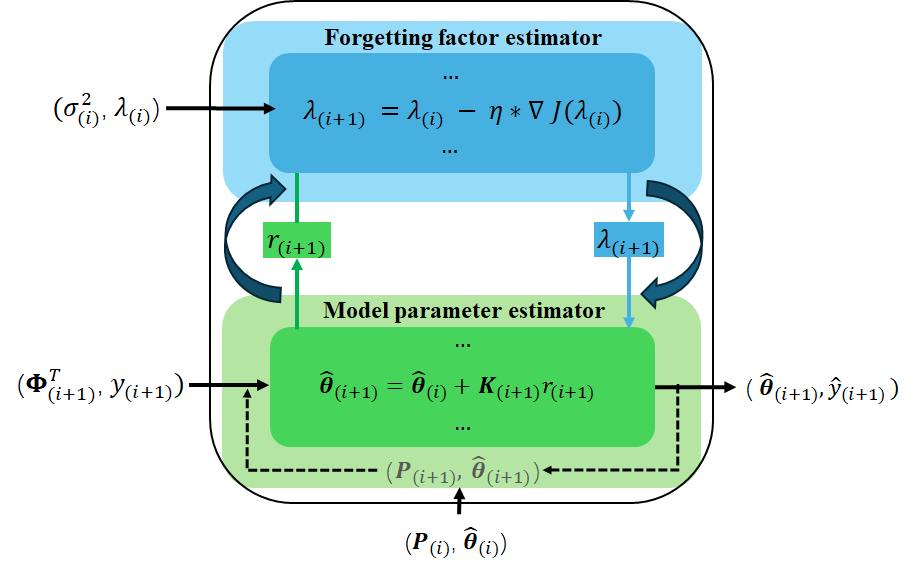}
    \caption{The architecture diagram of the RVM-RLS filter}
    \label{fig:GD_RLS_PR-label}
\end{figure}

Fig.~\ref{fig:GD_RLS_PR-label} shows the architecture of the RVM-RLS filter, which consists of the dual recursive estimation structure.
The first is a forgetting factor estimator that recursively updates the forgetting factor to track time-varying systems based on RVM paradigm using gradient descent [].
The second is the model parameter estimator (the proposed RLS framework in last subsection), which adaptively estimates the current parameter vector of the nonlinear system.
These two estimators operate in an alternating manner within a closed loop to achieve adaptive and accuracy estimation in time-varying nonlinear systems.

Specifically, the forgetting factor $\lambda_{(i+1)}$ is determined by minimizing a cost function $J(\lambda_{(i)})$ in the forgetting factor estimator.
Motivated by the principle that an estimator achieves optimality when the variance of the residuals matches the variance of the true noise, the residual variance matching estimation principle (RVME) is introduced (to be validated in next subsection). This paradigm formulates the cost function $J(\lambda_{(i)})$ by minimizing the variance discrepancy between the estimated residual variance $\hat{\sigma}_{(i+1)}^2$ and the true noise variance $\sigma^2$.

Finally, the cost function $J(\lambda_{(i)})$ is formulated in Equation~\eqref{GD-RLS-PR_lambda_form}:
\begin{equation}
	\label{GD-RLS-PR_lambda_form}
            \begin{aligned}
    J{(\lambda_{(i)})} &= c( (y_{(i+1)}- \hat{y}_{(i+1)} )^2 - {\sigma}^2)^2 \\
    &= c(r_{(i+1)}^2-{\sigma}^2)^2 \\
    &\approx c(\hat{\sigma}_{(i+1)}^2-{\sigma}^2)^2,
            \end{aligned}
\end{equation}
where $\hat{\sigma}{(i+1)}^2 = \lambda{(i)} \hat{\sigma}{(i)}^{2} + (1-\lambda{(i)}) r_{(i+1)}^{2}$ is formulated in a recursive structure to enable efficient online variance updates. In addition, $\hat{\sigma}{(i+1)}^2$ is adaptively constructed as a function of the variable forgetting factor $\lambda{(i)}$. $c$ is the proportion term
used to control the convergence speed. Accordingly, the gradient can be derived as shown in Equation~\eqref{GD-RLS-PR_lambda_gd}.

\begin{equation}
	\label{GD-RLS-PR_lambda_gd}
            \begin{aligned}
    \nabla J(\lambda_{(i)}) &
    =2c(\hat{\sigma}_{(i+1)}^2-\sigma^2)\cdot (\hat{\sigma}_{(i)}^2-r_{(i+1)}^2). \\
            \end{aligned}
\end{equation}
Therefore, forgetting factor $\lambda_{(i+1)}$ is expressed in \eqref{lambd_i+1} based on the gradient descent.
    \begin{equation}
	\label{lambd_i+1}
	\begin{split}
\lambda_{(i+1)} = \lambda_{(i)}- \eta \cdot \nabla J(\lambda_{(i)}),
	\end{split}
        \end{equation}
where $\eta$ denotes the step size used for updating. When the system changes rapidly, an excessively large forgetting factor $\lambda_{(i)}$ degrades the filter’s tracking performance; whereas when the system varies slowly, an overly small $\lambda_{(i)}$ may lead to overfitting. In both cases, the estimated residual variance increase which leads estimator accordingly adjusts $\lambda_{(i+1)}$ in the opposite direction to adapt and realize optimal performance adaptively.

Based on the updated forgetting factor $\lambda_{(i+1)}$, the proposed RLS framework (the model parameter estimator) subsequently updates the parameter vector $\mathbf{\hat{\boldsymbol{\theta}}}_{(i+1)}$ and proceeds to the next iteration.
So, the overall procedure of the proposed RVM-RLS filter is summarized in Algorithm~\ref{alg:AR-RLS}.

\begin{algorithm}
    {\setstretch{1.19}
	\caption{RVM-RLS Filter}\label{alg:AR-RLS} 
	\KwIn{$t_{(i)}, {{y}}_{(i)}$}
	\KwOut{$\boldsymbol{\hat{\theta}}_{i}, \hat{{y}}_{(i)}$}

    \KwInit{\\$(max(i)= N, m=4, M=i-m-1, \eta=0.001, c=20)$}\\
    \quad$\lambda_{(i)} = clip(\lambda_{min}=0.85,\lambda_{max}=0.95)$
    \\
    \quad $\hat{\sigma}^2=[(\mathbf{{y}}_{(i)}-\mathbf{\Phi}_{(i)}\boldsymbol{\hat{\theta}}_{(i)})^\top(\mathbf{{y}}_{(i)}-\mathbf{\Phi}_{(i)}\boldsymbol{\hat{\theta}}_{(i)})]/M$
    \\
    \quad$\boldsymbol{\hat{\theta}}_{(i)}=(\mathbf{\Phi}_{(i)}^\top\mathbf{\Phi}_{(i)})^{-1}\mathbf{\Phi}_{(i)}^\top\mathbf{{y}}_{(i)}$
    \\
    \quad$\mathbf{P}_{(i)}= \hat{\sigma}^2\cdot(\mathbf{\Phi}_{(i)}^\top\mathbf{\Phi}_{(i)})^{-1}$
    \\
    
	\While{$i$ $<$ $N$ }
    { 
    $\hat{y}_{(i+1)}=\mathbf{\Phi}_{(i)}^\mathsf{T} \hat{\boldsymbol{\theta}}_{(i)}$\\
    $r_{(i+1)}= y_{(i+1)}- \hat{y}_{(i+1)}$\\
    \If{${|r_{(i+1)}|>3}\sqrt{\hat{\sigma}^2}$}{$r_{(i+1)} = 0$\\
    break}
    $\hat{\sigma}_{(i+1)}^2 = \lambda_{(i)} \cdot\hat{\sigma}_{(i)}^{2} + (1-\lambda_{(i)}) \cdot r_{(i+1)}^{2}$   \\
    
    $J{(\lambda_{(i)})}  \approx c(\hat{\sigma}_{(i+1)}^2-{\sigma}^2)^2$\\
    
    $\nabla J(\lambda_{(i)}) = 2c(\hat{\sigma}_{(i+1)}^2-\sigma^2)\cdot (\hat{\sigma}_{(i)}^2-r_{(i+1)}^2)$\\
    
    $\lambda_{(i+1)} = clip (\lambda_{(i)}- \eta \cdot \nabla J(\lambda_{(i)}), \lambda_{min}, \lambda_{max})$ \\
    
    $W={\lambda_{_{(i+1)}} +\mathbf{\Phi}_{(i+1)}^T\mathbf{P}_{_{(i)}}\mathbf{\Phi}_{(i+1)}}$\\
   $\mathbf{K}_{(i+1)}=\frac{1}{W} \left( \mathbf{P}_{_{(i)}}\mathbf{\Phi}_{(i+1)} \right)$ \\
    $\hat{\boldsymbol{\theta}}_{(i+1)}=\hat{\boldsymbol{\theta}}_{(i)}+\mathbf{K}_{(i+1)} \cdot r_{(i+1)}$\\
    
    $\mathbf{P}_{(i+1)}=\frac{1}{\lambda_{(i+1)}}\left(\mathbf{P}_{(i)}-\mathbf{K}_{(i+1)}\mathbf{\Phi}_{(i+1)}^T\mathbf{P}_{(i)}\right)$\\
    $i= i+1$}
    }
\end{algorithm}

\subsection{MMSE Estimator and RVME Principle}
The introduced RVME principle enables the RVM-RLS filter to perform estimation by matching the residual variance to the noise variance, which is theoretically equivalent to the Minimum Mean Squared Error (MMSE) estimator under Gaussian noise, local linearization, and weak prior or consistent estimation conditions.

Consider the measurement model:
\begin{equation}
\mathbf{z} = \mathbf{h}(\mathbf{x}) + \mathbf{v}, \quad \mathbf{v} \sim \mathcal{N}(\mathbf{0}, \mathbf{R}),
\end{equation}
where $\mathbf{z}$ is the measurement from sensors, $\mathbf{x}$ is the true state, and $\mathbf{h}(\cdot)$ is the observation function. The residual is defined as $\mathbf{r} = \mathbf{z} - \mathbf{h}(\hat{\mathbf{x}})$ ($\hat{\mathbf{x}}$ is estimated state). $\mathbf{v}$ is the Gaussian noise, The $\mathbf{R}$ can be estimated or obtained.

Known that the true state can be estimated by the classical Minimum Mean Squared Error (MMSE) estimator:
\begin{equation}
\hat{\mathbf{x}}_{\text{(MMSE)}} = \arg\min_{\hat{\mathbf{x}}} \mathbb{E}\left[ \|\mathbf{x} - \hat{\mathbf{x}}\|^2 \mid \mathbf{z} \right].
\end{equation}
Under Gaussian assumptions and local linearization, the MMSE estimator satisfies the \textbf{orthogonality principle}:
\begin{equation}
\mathbb{E}\left[ \mathbf{r} (\mathbf{x} - \hat{\mathbf{x}})^T \right] = \mathbf{0}.
\end{equation}
This implies that the residual is \textbf{uncorrelated} with the estimation error.
The innovation covariance is:
\begin{equation}
\mathbf{S} = \mathbb{E}[\mathbf{r}\mathbf{r}^T] = \mathbf{H} \mathbf{P} \mathbf{H}^T + \mathbf{R},
\end{equation}
where $\mathbf{P}$ is the posterior covariance and $\mathbf{H}$ is the linearization of the $\mathbf{h}(\cdot)$
In the limit of weak prior or consistent estimation ($\mathbf{P} \to \mathbf{0}$), we have:
\begin{equation}
\mathbb{E}[\mathbf{r}\mathbf{r}^T] = \mathbf{R}
\end{equation}
Thus, \textbf{MMSE estimator enforces residual variance matching}.
At the same time, the introduced RVME principle minimizes:
\begin{equation}
\hat{\mathbf{x}}_{\text{(RVME)}} = \arg\min_{\hat{\mathbf{x}}} \left\| \mathbf{r}\mathbf{r}^T - \mathbf{R} \right\|_F^2.
\end{equation} This forces $\mathbf{r}\mathbf{r}^T = (\mathbf{z} - \mathbf{h}(\hat{\mathbf{x}}))(\mathbf{z} - \mathbf{h}(\hat{\mathbf{x}}))^T =\mathbf{R}$.
The optimization objectives between the MMSE and RVME are \textbf{identical} under the same assumptions. 

Therefore:
\begin{equation}
\hat{\mathbf{x}}_{\text{(MMSE)}} = \hat{\mathbf{x}}_{\text{(RVME)}}
\end{equation}

\section{The UAV-based Online terrain following System}
This section presents the design and modeling of a UAV-based online terrain following system for real-time waypoints generation and analyzes its uncertainty.

\subsection{System Design and Modeling}
\begin{figure}[h]
	\centering
	\includegraphics[width=3.0in]{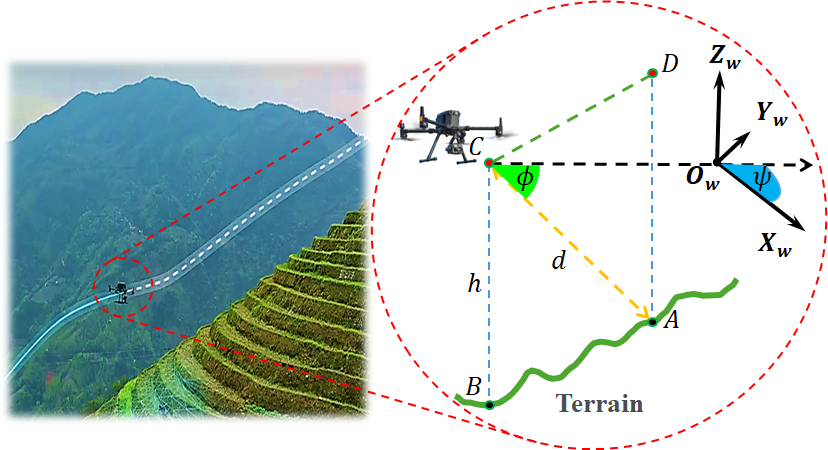}
	\caption{The diagram of the real-time waypoints generation based on UAV-based online terrain following system.}
	\label{fig_am}
\end{figure}

Fig.~\ref{fig_am} shows the diagram of the designed UAV-based online terrain following system.
The system consists of a UAV equipped with integrated navigation and control ability, and a gimbal-mounted LiDAR rangefinder. Through mutual collaboration, the UAV can generate and follow real-time path points during wildfire patrol. In the world coordinate system of $O_wX_wY_wZ_w$ determined by Real-Time Kinematic Global Positioning System (RTK-GPS), the pitch angle of the gimbal is $\phi$ in the longitudinal plane and the yaw angle of the UAV and gimbal all are $\psi$ in the horizontal plane.
During online terrain following, when the UAV reaches the current waypoint $C(x_c, y_c, z_c)$, the LiDAR rangefinder measures the reflection distance $d$ to point $A$ on the terrain surface. Based on the desired ground clearance $h$, which is determined by safety and mission requirements, the next real-time path point $D(x_d, y_d, z_d)$, which is located vertically above point $A$, is calculated in Eq.~\eqref{am} recursively.
\begin{equation}
	\label{am}
            \begin{aligned}
    x_d &= x_c + h \cdot\cos (\phi + v_{\phi})\cdot\cos(\psi) \\
    y_d &= y_c + h \cdot \cos (\phi + v_{\phi})\cdot\sin(\psi)\\
    z_{d} &= z_{c} + h - (d + v_{d})\cdot \sin{(\phi + v_{\phi})},
            \end{aligned}
\end{equation}
where, $v_d$ and $v_\phi$ denote the noise of the LiDAR rangefinder and gimbal, respectively. These errors are considered as independent Gaussian white noises, uncorrelated over time, with $v_d \sim \mathcal{N}(0, \sigma_{\mathrm{LiDAR}}^2)$ and $v_\phi \sim \mathcal{N}(0, \sigma_{\mathrm{gimbal}}^2)$.

\subsection{Uncertainty Analysis}

Given that vertical waypoints accuracy is crucial for UAV-based online terrain following, at the instant of sampling $t=i$, the next vertical waypoint $z_{(i+1)}$ is calculated in Eq.~\eqref{d_am2} based on Eq.~\eqref{am}
\begin{equation}
        \label{d_am2}
    z_{(i+1)} = z_{(i)} + h - (d_{(i)} + v_{d(i)})\cdot \sin{(\phi_{(i)} + v_{\phi(i)})},
\end{equation}
where, $z_{(i)}$, $d_{(i)} + v_{d(i)}$ and $\phi_{(i)} + v_{\phi(i)}$ are observation sequences of the RTK-GPS positioning, distance of the rangefinder and the pitch angle of the gimbal, respectively.
Therefore, series $z_{(i)}$ is regarded as an equivalent sensor observation sequence, synthesized from multiple sensors including RTK-GPS, LiDAR Rangefinder and gimbal.
Consequently, the observation $z_{(i+1)}$ can serve as the input of the AR-RLS for subsequent filtering.

without considering the positioning error introduced the RTK-GPS (RTK-GPS can typically provide sub-centimeter accuracy) and according to the propagation of uncertainty, the uncertainty of this system can be evaluated and represented by the variance $\sigma_{z}^2$ in \eqref{all_am}:
\begin{equation}
    \label{all_am}
    \sigma_{\Delta z}^2\approx\left(\frac{\partial \Delta z}{\partial v_{d}}\right)^2 \cdot \sigma_{\mathrm{LiDAR}}^2+\left(\frac{\partial \Delta z}{\partial v_{\phi}}\right)^2 \cdot \sigma_{\mathrm{gimbal}}^2,
\end{equation}
where, the partial derivative values are expressed in~\eqref{pd_am}:
\begin{equation}
	\label{pd_am}
            \begin{aligned}
    \frac{\partial {\Delta z}}{\partial v_d} &\approx-\sin(\phi + v_\phi) \\
    \frac{\partial {\Delta z}}{\partial v_{\phi}} &\approx -(d + v_d) \cdot \cos(\phi + v_\phi).
            \end{aligned}
\end{equation}
When $v_{d}\approx0,v_{\mathrm{\phi}}\approx0$, the standard deviation of the $\Delta z$ is estimated as \eqref{totall_variance}
\begin{equation}
    \label{totall_variance}
\sigma_{\Delta z}=\sqrt{\sin^2(\phi)\cdot\sigma_{\mathrm{LiDAR}}^2 + d^2\cdot\cos^2(\phi)\cdot\sigma_{\mathrm{gimbal}}^2} ,
\end{equation}

Therefore, the uncertainty of this online terrain following waypoints depends not only on the $\phi$ and $v_\phi$, but also directly on the $d$ and $v_d$, which can significantly amplify overall positioning error. 

\section{simulation and results}
\begin{figure*}[!h]
    \centering
    \includegraphics[width=0.95\textwidth]{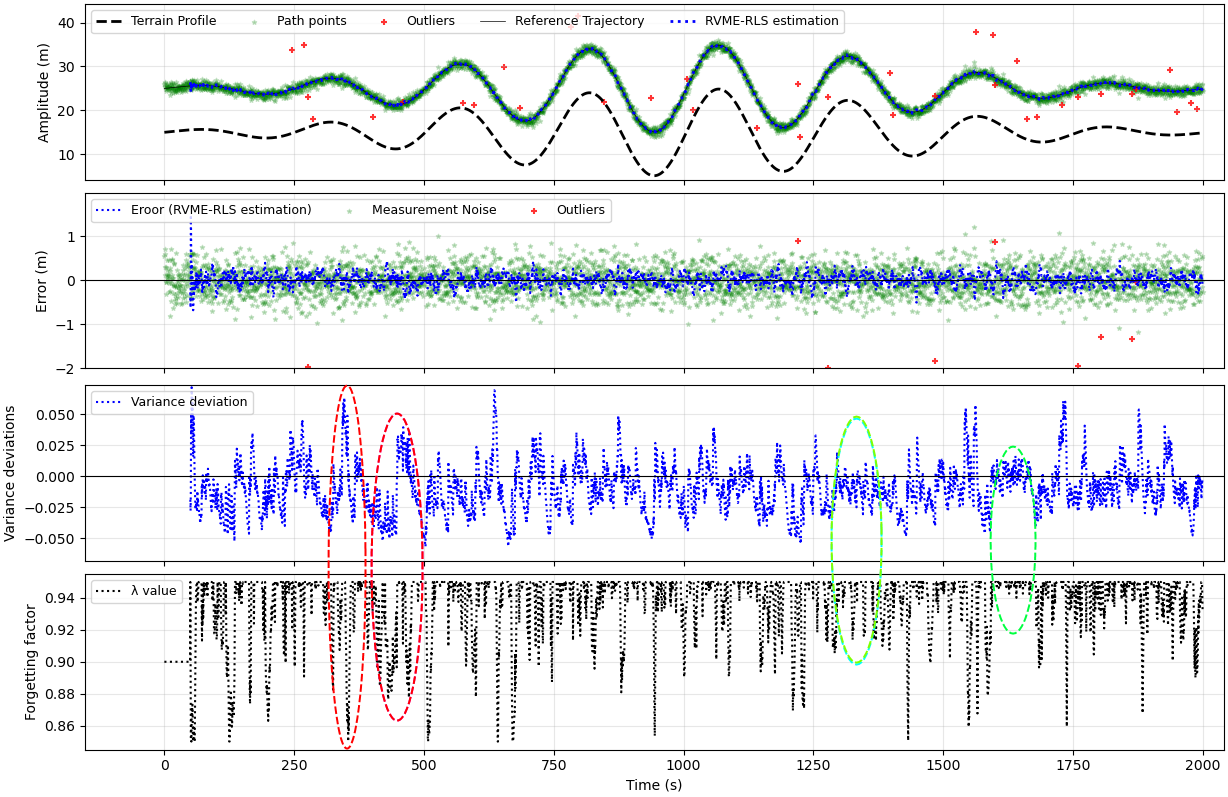}
    \caption{The simulation results of the Residual PolyRLS filter.}
    \label{fig: simulation_gd-rls}
\end{figure*}
This section presents the simulation setup, performance evaluation, and comparative analysis conducted to verify the effectiveness of the proposed RVM-RLS filter.

\subsection{Simulation Setup}

Real-time vertical waypoint estimation is simulated for the UAV-based terrain-following system under random disturbances (noise and outliers) in a nonlinear and dynamic terrain environment.
As illustrated in the first subfigure of Fig.~\ref{fig: simulation_gd-rls}, the terrain profile follows a sinusoidal pattern with a time-varying amplitude, representing nonlinear and dynamic characteristics. Its mathematical expression is defined as

\begin{equation}
	\label{terrain_profile}
            \begin{aligned}
    H_{(t)}=A(t)\sin(\omega_\mathrm{ter~}t+\phi_\mathrm{ter}),
            \end{aligned}
\end{equation}
where, $A(t)=10\mathrm{~exp}\left[-\frac{(t-1000)^2}{2\sigma_{ter}^2}\right](\sigma_{ter}=400)$, is the Gaussian term which controls the amplitude envelope, producing a bell-shaped modulation centered at $t=1000$ and the sinusoidal term $sin(\omega_\mathrm{terrain~}t+\phi_\mathrm{ter})(\omega_\mathrm{ter}=0.025,\phi_\mathrm{ter}=0)$, represents the oscillatory component of the terrain profile.

Besides, the reference trajectory for terrain following is defined as Eq.~\eqref{H_t}
\begin{equation}
	\label{H_t}
            \begin{aligned}
    p_{(t)}=H_{(t)} + h,
            \end{aligned}
\end{equation}
where $h$ denotes the desired flight altitude above the terrain.
The green star points represent the real-time path points computed by the online UAV terrain-following system under environmental disturbances, which can be modeled as  Eq.~\eqref{z_t}
\begin{equation}
	\label{z_t}
            \begin{aligned}
    z_{(t)}=p_{(t)}+v_{(t)}+o_{(t)}+s_{(t)},
            \end{aligned}
\end{equation}
where the term $v_{(t)}$ denotes the measurement noise of this system, modeled as a zero-mean Gaussian process $v_{(t)} \sim \mathcal{N}(0, \sigma^2)$. The term $o_{(t)}$ represents outliers (red cross points) arising from sensor faults, data conversion errors, and environmental disturbances occurring at random and unknown instants. These outliers significantly deviate from normal observations, typically manifesting as abrupt spikes or drops in the data. 
In this simulation, the proportion of outliers is constrained to 10$\%$ of the total number of path points, and their impulsive amplitude are constrained within the range [$-30\sigma$,30$\sigma$].
The term $s_{(t)}$ denotes other error sources that are neglected in this study.

\subsection{Performance of the RVM-RLS Filter}
\begin{figure*}[h!t]
	\centering
\includegraphics[width=0.95\textwidth]{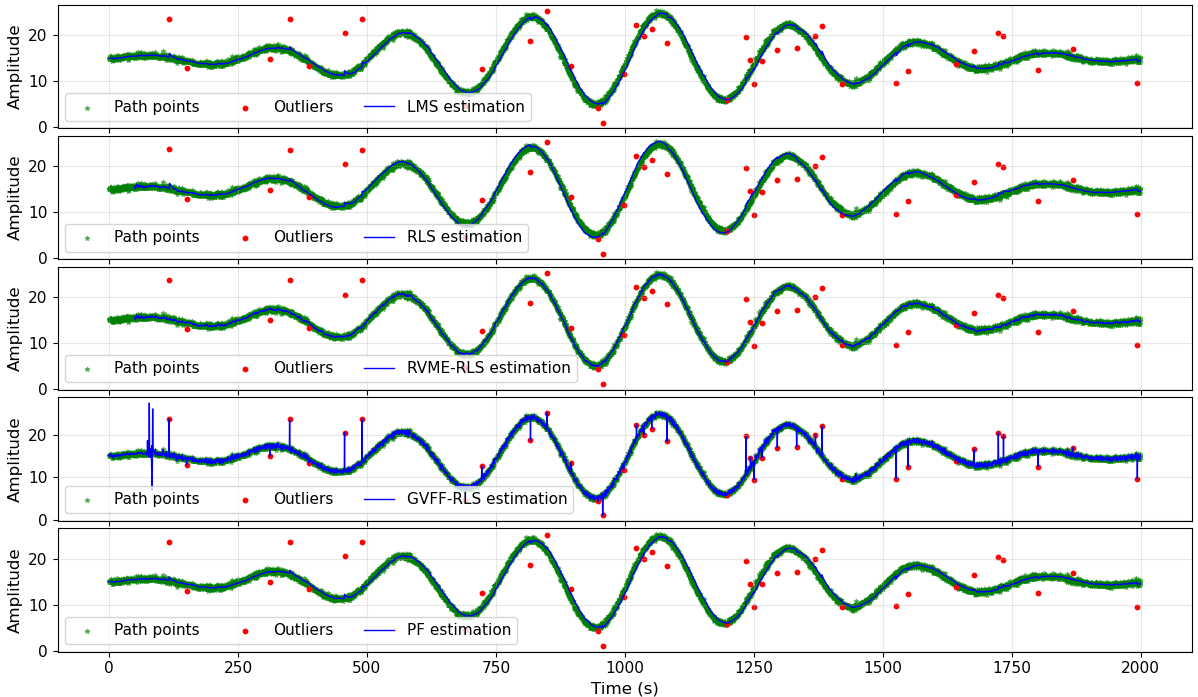}
	\caption{Comparative simulation of estimated waypoints during online terrain following.}
    \label{ACV2}
\end{figure*}
This subsection evaluates the performance of the proposed RVM-RLS filter in processing real-time waypoints contaminated by random noise and outliers.
As shown in the first subfigure of Fig.~\ref{fig: simulation_gd-rls}, the blue dotted curve represents the estimated path points generated by the RVM-RLS filter, which effectively suppresses both noise and outliers. The second subfigure presents the path point noise $v_{(t)}$, outliers $o_{(t)}$, and the residuals of the RVM-RLS filter (i.e., the deviation between the RVM-RLS curve and the reference trajectory). It can be observed that the residuals exhibit significantly improved precision, confirming the filter’s ability to mitigate measurement disturbances.

The third subfigure illustrates the temporal evolution of the variance deviation, defined as the sequence of differences between the estimated and true noise variances, while the fourth subfigure shows the corresponding adaptive forgetting factor. A closer examination reveals an inverse relationship between the variance deviation and the forgetting factor: larger variance deviations correspond to smaller forgetting factors (indicated by red ellipses), whereas smaller deviations correspond to larger forgetting factors (indicated by green ellipses).

Integrating the observations across all subfigures of Fig.~\ref{fig: simulation_gd-rls}, the results confirm that the RVM-RLS filter effectively suppresses measurement noise and outliers while accurately tracking the nonlinear dynamics of the time-varying system. Furthermore, the observed correlation between the variance deviation and the forgetting factor demonstrates that the proposed RVM-RLS filter adaptively adjusts its forgetting factor in response to changes in variance deviation. This adaptive adjustment behavior aligns with the theoretical design of the RVM-RLS framework, validating its capability to achieve high-accuracy, adaptive real-time state estimation under nonlinear and time-varying conditions.

\subsection{Comparison Simulation and Error Analysis}
To comprehensively evaluate the performance of the proposed RVM-RLS filter, comparative simulations are conducted against four representative filtering algorithms, including the LMS, the RLS, the Gradient-based Variable Forgetting Factor RLS (GVFF-RLS) \cite{leung2005gradient} and the PF.
To assess performance quantitatively, four metrics are employed: Single-step Runtime (SR), Mean Squared Error (MSE) \cite{wasilewska2025protection}, Variance Ratio (VR), and Maximum Error (ME).
All simulations were performed on a laptop equipped with an AMD Ryzen 7 5800U CPU (8 cores, 1.90 GHz base), 16 GB DDR5 RAM, running Windows 11 64-bit.
The algorithm was implemented in Python 3.12 using NumPy 1.26.4 and SciPy 1.15.2 without GPU acceleration.
A smaller MSE reflects higher estimation accuracy and convergence stability.  
The ME measures the algorithm’s robustness to outliers. The SR evaluates computational efficiency, which is particularly critical for real-time applications.
The VR, defined in Eq.~\eqref{VR}, quantifies the measurement noise suppression capability by comparing the variance before and after filtering:
\begin{equation}
	\label{VR}
            \begin{aligned}
    VR=\frac{\sigma_r^2}{{\sigma^2}},
            \end{aligned}
\end{equation}
where $\sigma_r^2$ denotes the error variance after filtering, and $\sigma^2$ is the measurement noise variance before filtering. The VR value closer to zero indicates stronger measurement noise suppression ability.

\begin{figure*}[h!t]
	\centering
\includegraphics[width=0.95\textwidth]{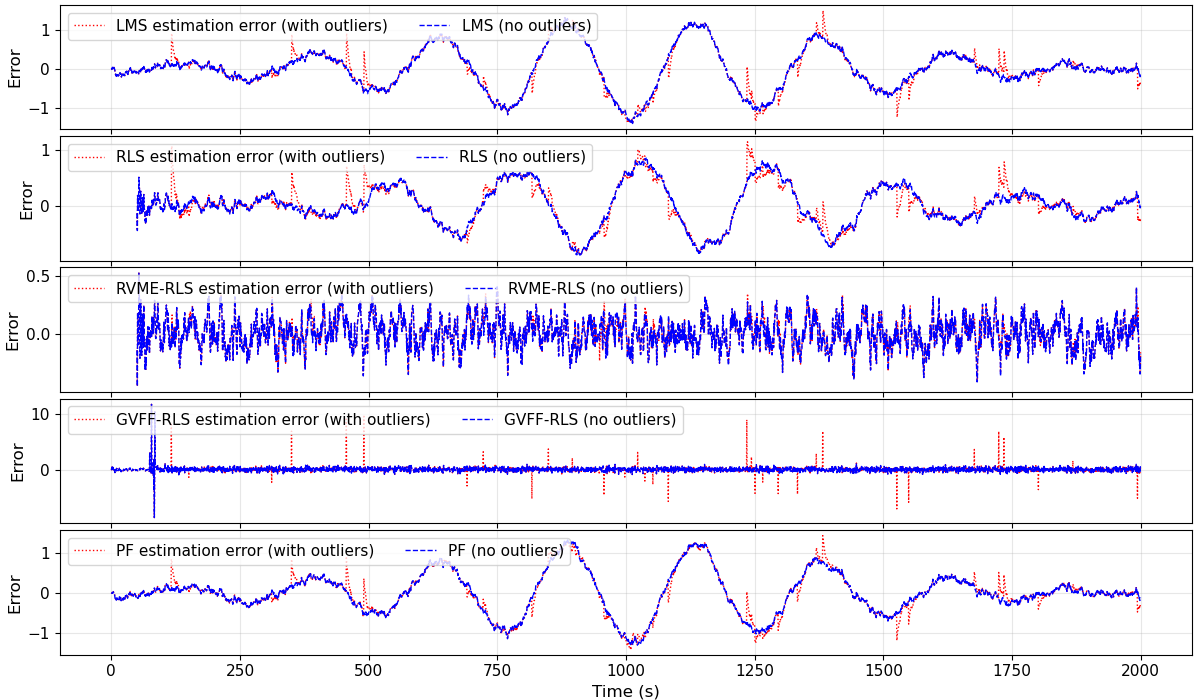}
	\caption{Comparison of estimation errors among different algorithms during online terrain-following simulation.}
    \label{EE}
\end{figure*}

All of the relevant algorithms are implemented to process the path points $z_{(t)}$ ($\sigma^2 = 0.09$). Fig.~\ref{ACV2} separately presents the corresponding estimated path points of different algorithms and the path points $z_{(t)}$.  
The corresponding algorithms estimation error defined as the deviation between the estimated path points and the reference trajectory $p_{(t)}$ are shown separately in Fig.~\ref{EE} to highlight the detailed.
It can be observed that, in Fig.~\ref{ACV2} and Fig.~\ref{EE}, the GVFF-RLS algorithm exhibits poor robustness against outliers and fails to adapt to the nonlinear terrain variations. The LMS, RLS, and PF algorithms achieve only moderate filtering performance, providing limited suppression of outlier disturbances and insufficient tracking accuracy under nonlinear dynamics.
In contrast, the proposed RVM-RLS filter demonstrates superior robustness and adaptability, effectively rejecting outliers while accurately tracking nonlinear terrain variations, thereby achieving stable and high-precision real-time estimation.

Tables~\ref{tab:algorithm_PP_p} and~\ref{tab:algorithm_PP_1} summarize the quantitative performance of all algorithms under outlier and non-outlier conditions, respectively. The proposed RVM-RLS filter exhibits complete immunity to outliers and achieves the highest overall filtering accuracy. In contrast, the LMS, PF, and RLS algorithms show limited robustness against outliers, while the GVFF-RLS fails entirely in their presence.
Although the LMS and RLS algorithms demonstrate excellent computational efficiency, their robustness and estimation precision are markedly inferior to those of the RVM-RLS, regardless of outlier interference. Both GVFF-RLS and RVM-RLS maintain acceptable real-time performance for online terrain-following applications, whereas the PF algorithm suffers from excessive computational cost due to its particle-based iterative process, making it unsuitable for real-time implementation.
These quantitative results are consistent with the trends observed in Fig.~\ref{ACV2} and~\ref{EE}, further confirming the superior performance of the proposed RVM-RLS algorithm. Based on the MSE and VR metrics, the RVM-RLS improves filtering accuracy by approximately 88$\%$ compared with the conventional RLS under outlier conditions.
In summary, the proposed RVM-RLS consistently achieves an optimal balance among accuracy, robustness, and computational efficiency, demonstrating strong adaptability to measurement uncertainties and nonlinear time-varying systems.
\begin{table}[h]
\caption{Algorithms Performance Comparison  without outliers}
\begin{center}
\begin{tabular}{|c|c|c|c|c|}
\hline
\textbf{Algorithm} & \multicolumn{4}{|c|}{\textbf{Performance Indices} 
  [$\sigma^2=0.09 (m^2)$]} \\
\cline{2-5} 
 & \textbf{\textit{SR (ms)}} & \textbf{\textit{MSE}} & \textbf{\textit{VR}} & \textbf{\textit{ME}} \\
\hline
LMS & 0.005 & 0.280 & 0.311 & 1.397  \\
\hline
RLS & 0.014 & 0.128 & 1.424 & 0.893  \\
\hline
RVM-RLS & \textbf{0.024} & \textbf{0.015} & \textbf{0.172} & \textbf{0.524} \\
\hline
GVFF-RLS & 0.023 & 0.180 & 1.995 & 11.830  \\
\hline
PF & 0.765 & 0.268 & 2.977 & 1.341  \\
\hline
\end{tabular}
\label{tab:algorithm_PP_p}
\end{center}
\end{table}

\begin{table}[h]
\caption{Algorithms Performance Comparison with outliers}
\begin{center}
\begin{tabular}{|c|c|c|c|c|}
\hline
\textbf{Algorithm} & \multicolumn{4}{|c|}{\textbf{Performance Indices} 
  [$\sigma^2=0.09 (m^2)$]} \\
\cline{2-5} 
 & \textbf{\textit{SR (ms)}} & \textbf{\textit{MSE}} & \textbf{\textit{VR}} & \textbf{\textit{ME}}  \\
\hline
LMS & 0.004 & 0.295 & 3.273 & 1.514 \\
\hline
RLS & 0.014 & 0.132 & 1.467 & 1.154  \\
\hline
RVM-RLS & \textbf{0.024} & \textbf{0.016} & \textbf{0.173} & \textbf{0.524} \\
\hline
GVFF-RLS & 0.023 & 0.401 & 4.457 & 11.830  \\
\hline
PF & 0.765 & 0.278 & 3.089 & 1.441  \\
\hline
\end{tabular}
\label{tab:algorithm_PP_1}
\end{center}
\end{table}

\section{CONCLUSIONS}
This paper presents the Residual Variance Matching–Recursive Least Squares (RVM-RLS) filter, a novel adaptive and robust estimator for real-time waypoints estimation in UAV-based online terrain following, robust to outliers. To evaluate the approach, we developed a dedicated UAV-based online terrain-following system paired with a simulated terrain environment to generate raw path points in real-time.

Based on across multiple metrics, the performance of the RVM-RLS filter was rigorously compared against several benchmark algorithms . Simulation results demonstrate its superior estimation accuracy, enhanced outlier rejection, and excellent adaptability to nonlinear, time-varying systems. Quantitatively, the proposed filter achieves an approximately 88$\%$ improvement in path point estimation accuracy over the standard RLS, as measured by mean squared error (MSE) and variance ratio (VR), while maintaining acceptable real-time performance.

Despite these advances, the method remains constrained by its static noise variance assumption and underutilizes the rich information embedded in real-time measurements. 
Future work will extend the RVM-RLS framework to incorporate dynamic noise variance estimation and data fusion, further enhancing its adaptation and state estimation capabilities in complex and uncertain environments.

\addtolength{\textheight}{-12cm}   



\section{ACKNOWLEDGMENT}
This work was supported in part by the Natural Sciences and Engineering Research Council of Canada (NSERC) through a Discovery project and the CRIAQ through the project Système de Guidage par Vision Intelligente de Poursuite de Trajectoire sur Aéronefs Bombardiers d'Eau $/$ AirTanker Visual Intelligent Tracking \& Airborne Guidance System (AVITAGS).

\end{document}